\documentclass[epj]{webofc}
\usepackage[utf8]{inputenc}
\usepackage[varg]{txfonts}   
\usepackage{booktabs}
\usepackage{xcolor}
\definecolor{darkred}{rgb}{0.4,0.0,0.0}
\definecolor{darkgreen}{rgb}{0.0,0.4,0.0}
\definecolor{darkblue}{rgb}{0.0,0.0,0.4}
\usepackage[bookmarks,linktocpage,colorlinks,
    linkcolor = darkred,
    urlcolor  = darkblue,
    citecolor = darkgreen]{hyperref}
\usepackage{subfigure}
\wocname{EPJ Web of Conferences}
\woctitle{Lattice2017}
\begin{document}
%
\selectlanguage{english}
\title{%
Can a Linear Sigma Model Describe Walking Gauge Theories at Low Energies?
}
\author{%
\firstname{Andrew} \lastname{Gasbarro}\inst{1,2,3}\fnsep\thanks{Speaker, \email{andrew.gasbarro@yale.edu}.  \textbf{Acknowledgments: }This work is carried out as a part of the Lattice Strong Dynamics Collaboration (LSD).  This material is based upon work supported by the U.S. Department of Energy, Office of Science, Office of Workforce Development for Teachers and Scientists, Office of Science Graduate Student Research (SCGSR) program.  The SCGSR program is administered by the Oak Ridge Institute for Science and Education (ORISE) for the DOE.  ORISE is managed by ORAU under contract number DE-SC0014664.  A.G. thanks Lawrence Livermore National Laboratory and Lawrence Berkeley National Laboratory for hospitality during the completion of this work.  He also thanks James Ingoldby and LSD members George Fleming, Thomas Appelquist, Pavlos Vranas, and Richard Brower for many useful discussions.}
}
\institute{%
Sloane Physics Laboratory
Yale University
New Haven, CT 06520, USA
\and
Lawrence Livermore National Laboratory, Livermore, CA 94550, USA
\and
Lawrence Berkeley National Laboratory, Berkeley, CA 94720, USA
}

\abstract{%
  In recent years, many investigations of confining Yang Mills gauge theories near the edge of the conformal window have been carried out using lattice techniques. These studies have revealed that the spectrum of hadrons in nearly conformal ("walking") gauge theories differs significantly from the QCD spectrum. In particular, a light flavor-singlet scalar appears in the spectrum which is nearly degenerate with the pseudo-Nambu-Goldstone bosons (PNGBs) at the lightest currently accessible quark masses. This state is a viable candidate for a composite Higgs boson.
  Presently, an acceptable effective field theory (EFT) description of the light states in walking theories has not been established. Such an EFT would be useful for performing chiral extrapolations of lattice data and for serving as a bridge between lattice calculations and phenomenology. It has been shown that the chiral Lagrangian fails to describe the IR dynamics of a theory near the edge of the conformal window. 
  Here we assess a linear sigma model as an alternate EFT description by performing explicit chiral fits to lattice data.  
  Our model provides an acceptable combined fit to the PNGB mass and decay constant, reducing the $\chi^2$/d.o.f. by more than an order of magnitude compared to next-to-leading order chiral perturbation theory.  
}
\maketitle
\section{Introduction}

In recent years, there has been a renewed interest in the capacity of strongly coupled Yang-Mills gauge theories to admit low energy physics that is markedly different from the low energy behavior of Quantum Chromodynamics (QCD).  
The novel physics seems to appear concurrently with the theory becoming nearly scale invariant in the IR in a region of parameter space known as the conformal window.  
Currently, the primary tool to study theories near and inside the conformal window is lattice Monte Carlo calculation.  
However, in a scale invariant system correlations quickly grow to be as large as the box size and finite volume effects become problematic.  
A new lattice methodology capable of studying scale invariant theories in radial quantization on the lattice could alleviate these problems in the future \cite{Brower:2012vg,Brower:2016moq,Brower:2016vsl}, but these techniques are still being developed for four dimensions.
We are left to work with data from standard lattice methods for now.  
In lattice QCD, a similar problem arises when one tries to take the chiral limit.  
Goldstone modes approach an infinite correlation length, and the box size must be taken ever larger as the quark masses are brought to zero.  
The chiral limit is never truly reached; instead, one typically uses an effective field theory (EFT) -- often chiral perturbation theory ($\chi$PT) -- to extrapolate to the chiral limit.  
In a nearly scale invariant theory the problem is compounded because as the quark masses are taken to zero not only do the Goldstone masses become small compared to the confinement scale but the confinement scale becomes small compared to the lattice cutoff.
An EFT description that is reliable at currently reachable distances from the chiral limit would greatly aid in understanding these theories.

We are ultimately interested in scenarios in which electroweak symmetry is spontaneously broken by a new strong force, so in this work we focus on \emph{walking theories} outside the edge of the conformal window, which have a small but non-vanishing beta function and spontaneously broken chiral symmetry in the IR.  
SU$(3)$ gauge theory with $N_f = 8$ flavors of quarks in the fundamental representation has been shown to have a small but nonzero beta function \cite{Hasenfratz:2014rna,Fodor:2015baa}, which makes it a good exemplar for understanding walking theories.  
Studies of the spectrum of the $N_f = 8$ theory \cite{Aoki:2014oha,Appelquist:2016viq,rinaldi} have revealed that the $\sigma$ meson 
\footnote{Throughout this work, we borrow the language of QCD for labeling hadron states.  For the lightest resonance in each channel, we denote the flavor-singlet scalar by $\sigma$, the flavor-singlet pseudoscalar by $\eta'$,  the flavor-adjoint scalar by $a_0$, the flavor-adjoint pseudoscalar by $\pi$, the flavor-adjoint vector by $\rho$, and spin-1/2 baryon by $N$ or ``nucleon''. }
 is nearly degenerate with the pions and significantly lighter than the $\rho$ over a wide range of bare quark masses studied, which marks a substantial deviation from the more familiar spectrum of QCD.  
 In QCD, the $\sigma$ is several times heavier than the pions and unstable at a comparable distance from the chiral limit \cite{Briceno:2016mjc}.
 Since there is no gap between the pions and the $\sigma$ it is unlikely the $\chi$PT is applicable to the $N_f=8$ theory in the current regime of $m_q$.  
 A direct attempt at fitting $\chi$PT expressions to Goldstone observables in this theory \cite{Gasbarro:2017fmi} confirmed that $\chi$PT is not a good description of the LSD collaboration data.  
SU$(3)$ gauge theory with two flavors in the sextet representation -- also believed to exhibit walking dynamics -- has been shown to possess its own light $0^{++}$ state in its spectrum \cite{Fodor:2016pls}, which indicates that the light $\sigma$ may be a generic feature of walking theories.
Some expect that the $\sigma$ state in walking theories is a pseudodilaton whose small mass results from the softly broken scale invariance \cite{Yamawaki:1985zg,Appelquist:2010gy}, though this is far from proven.  
There have been efforts to develop an EFT based on the hypothesis of softly broken scale symmetry \cite{Matsuzaki:2013eva,Golterman:2016lsd,Appelquist:2017wcg}, but we take a different approach here.

In this note, we consider a generalized linear sigma model EFT in which the pions and light $\sigma$ transform together in a linear multiplet.
In the development of QCD, both the linear and nonlinear sigma models were considered as effective descriptions of the light states (cf \cite{weinberg}).
Since the QCD $\sigma$ particle is heavy and unstable, it was realized that integrating out this mode led to a more reliable effective theory for most applications.
For the $N_f = 8$ theory, since the $\sigma$ is light and stable at comparable distances from the chiral limit, it seems natural to reconsider the linear sigma model as an effective description.  
We also remark that this work is a step towards connecting the lattice calculations of walking theories with the phenomenology of electroweak symmetry breaking.  In the Standard Model, the Higgs sector is a linear sigma model, so agreements or deviations between the lattice data of the walking theory and the linear sigma model EFT begin to speak to the phenomenology of these gauge theories.

\section{Linear $\sigma$ Model for General $N_f$}

Here we present a construction of a generalized linear sigma model EFT appropriate for application to walking gauge theories.  
We will omit the flavor singlet pseudoscalar $\eta'$ degree of freedom whose mass is large due to the anomaly.  
$M_{\eta'}$ has been recently computed on the lattice by the latKMI collaboration and found to be much heavier than the $\rho$ in the $N_f = 8$ theory \cite{rinaldi}.  
We will also omit determinant operators related to the anomaly as we consider them to be higher dimensional in the power counting.  
For more details on the possible role of the anomaly in the linear sigma model applied to walking theories, see the recent work \cite{Meurice:2017zng}.

\subsection{Field Content}
The primary ansatz of the linear sigma EFT construction is that the fields parameterizing the low energy degrees of freedom transform in a linear multiplet of the global symmetry group.  
Similar constructions have been carried out for three flavor QCD \cite{Schechter:1993tc}.
The underlying walking gauge theory has the global symmetry U$_L(N_f)\times$U$_R(N_f)$ at the classical level.  
The U$_A(1)$ symmetry is anomalously broken by topology in the gauge fields, so we omit it.  
The U$_V(1)$ baryon number symmetry is trivial in a theory of only mesons. 
Therefore, we consider a matrix field transforming linearly under SU$_L(N_f) \times $SU$_R(N_f)$.  
Written out explicitly, the dynamical fields carry two indices, $M_a^{\bar{b}}(x)$, where the unbarred subscript (barred superscript) transforms via linear action of a matrix in the fundamental (antifundamental) representation of SU$_L(N_f)$  (SU$_R(N_f)$).
\begin{equation}
M_a^{\bar{b}} \rightarrow L_a^c M_c^{\bar{d}} \left(R^\dagger \right)_{\bar{d}}^{\bar{b}}
\end{equation}
where $L,R \in $ SU$_{L,R}(N_F)$.  Indices will be suppressed in the remainder of the discussion.  We choose to parametrize the degrees of freedom in a linear basis.
\begin{equation}
M(x) = \frac{\sigma(x)}{\sqrt{N_f}} + a_a(x)T^a + i \pi_x(x) T^a \label{eq:linearparam}
\end{equation}
$T^a$ are the generators of SU$(N_f)$ normalized such that $\langle T^a T^b \rangle = \delta^{ab}$ where $\langle ... \rangle$ denotes the trace.  The degrees of freedom are real scalar and pseudoscalar fields.  We have omitted the imaginary part of the trace, which is the $\eta'$ degree of freedom.  Notice that the linear multiplet includes not only the flavor singlet scalar ($\sigma$) and flavor adjoint pseudoscalar ($\pi_a$) degrees of freedom, but also an adjoint multiplet of scalars ($a_a$).  These states must be included because for $N_f \neq 2$ the linear representation of the group is necessarily complex.  For the special case of $N_f = 2$, the isometry SU$(2)\times$SU$(2) \sim O(4)$ allows for a real linear representation, which is equivalent to setting $a_a(x) = 0$ in Eq.~\ref{eq:linearparam}.  In what follows, we consider only the tree level expressions for the masses and decay constants of the $\sigma$ and $\pi$ states, and the flavored scalars play no role in these expressions.  A thorough discussion of the flavored scalar states is left to an upcoming publication \cite{Browerprogress}.

\subsection{Leading Order}
The leading order Lagrangian is given by 
\begin{flalign}
	\mathcal{L}_{LO} &= \frac{1}{2} \left\langle \partial_\mu M \partial^\mu M^\dagger \right\rangle - V_0(M) - V_{SB}(M,\chi) \label{eq:LLO} \\
	V_0 &= \frac{-m_{\sigma}^2}{4} \left\langle M^\dagger M \right\rangle + \frac{m_{\sigma}^2 - m_a^2}{8 f^2} \left \langle M^\dagger M \right\rangle^2 + \frac{N_f m_a^2}{8 f^2} \left \langle \left ( M^\dagger M \right)^2 \right\rangle \label{eq:v0}
\end{flalign}
where the unbroken potential $V_0$ contains all relevant and marginal operators invariant under the global symmetries. 
The breaking potential $V_{SB}$ depends on the choice of power counting as detailed in the next section.  
We have chosen to parametrize the couplings in the unbroken potential such that the Lagrangian takes a simple form after spontaneous symmetry breaking when $V_{SB} = 0$.  
When the symmetry breaks -- either spontaneously or explicitly -- the field acquires a vacuum expectation value (v.e.v.) which we choose by convention to be oriented along the direction of the trace (or the ``$\sigma$ direction'').  The minimum is given by $\sigma = F$, $a_a = \pi_a = 0$ where $F$ is determined by solving
\begin{equation}
\frac{F^2}{f^2} - 1 + \frac{2}{m_{\sigma}^2 F} \left. \frac{\partial V_{SB}}{\partial \sigma} \right|_{\sigma = F,\; a_a=\pi_a = 0} = 0 \label{eq:minimum}
\end{equation}
Re-expanding around this v.e.v. leads to the following leading order expressions for the masses of states.
\begin{flalign}
	M_{\pi}^2 &= \left.\frac{\partial^2 V_{\text{SB}}}{\partial \pi_a^2} \right|_{\sigma = F,\; a_a=\pi_a = 0} \label{eq:LO1} \\
	M_{\sigma}^2 &= m_\sigma^2 \left( \frac{3}{2} \frac{F^2}{f^2} - \frac{1}{2} \right) + \left.\frac{\partial^2 V_{\text{SB}}}{\partial \sigma^2}\right|_{\sigma = F,\; a_a=\pi_a = 0} \label{eq:LO2} \\
	M_{a}^2 &= m_a^2 \frac{F^2}{f^2} + \frac{m_{\sigma}^2}{2} \left(  \frac{F^2}{f^2} -1 \right) + \left.\frac{\partial^2 V_{\text{SB}}}{\partial a_a^2}\right|_{\sigma = F,\; a_a=\pi_a = 0} \label{eq:LO3}
\end{flalign}
In the absence of a breaking potential, $F = f$, $M_\sigma = m_\sigma$, $M_a = m_a$, and $M_\pi = 0$.  The parameters in $V_0$ are simply the chiral limit values of the masses and v.e.v.  

There is only one dimensionful quantity to serve as a scale in the chiral limit theory, the v.e.v. of the $\sigma$ field, $f$.  Accordingly, the scales of spontaneous chiral symmetry breaking and scale symmetry breaking are related in a linear sigma model.  Let us examine the PCAC relation for the leading order Lagrangian of Eqs.~\ref{eq:LLO},\ref{eq:v0}.  We follow normalization conventions for the external axial current and the PCAC matrix element given in \cite{Pich:1995bw}.  The leading order expression for the axial current is given by
\begin{equation}
A^{a\mu}(x) = \frac{i}{2} \left\langle \left\{T^a,M\right\}\partial^\mu M^\dagger - \left\{T^a,M^\dagger \right\}\partial^\mu M \right\rangle \rightarrow \frac{2 F}{\sqrt{N_f}} \partial^\mu \pi^a + ... \label{eq:axial}
\end{equation}
In the rightmost expression we have expanded about the scalar v.e.v. and written out only the term linear in the pion field that contributes to the PCAC matrix element.  The pion decay constant is defined as 
\begin{equation}
\left\langle 0  \left| A^{a\mu}(x) \right| \pi^b(\vec{p}) \right\rangle = i \sqrt{2} F_\pi p^\mu \delta^{ab} e^{ipx} \label{eq:pcac}
\end{equation}
Plugging Eq.~\ref{eq:axial} into Eq.~\ref{eq:pcac}, one finds $F_\pi = \sqrt{2/N_f} F$.  

Notice the important feature of the linear sigma model that $F_\pi$ inherits a tree level dependence on the breaking potential through Eq.~\ref{eq:minimum}.  In Ref.~\cite{Gasbarro:2017fmi}, the difficulty in fitting $\chi$PT expressions to the $N_f = 8$ theory was that $M_\pi^2$ favored small NLO corrections while $F_\pi$ favored large NLO corrections.  $M_\pi^2$ is highly linear in the quark mass in accordance with leading order $\chi$PT, but $F_\pi$ varies significantly with the quark mass in contrast with the leading order $\chi$PT prediction that $F_\pi$ is a constant.  The linear sigma model stands to alleviate this issue by virtue of the nontrivial leading order behavior of $F_\pi$.  We examine this by direct fits in Section \ref{sec:fits}.  As $m_\sigma$ becomes large, the tree level dependence of $F_\pi$ on $V_{SB}$ through Eq.~\ref{eq:minimum} is reduced.  It is precisely the lightness of the $\sigma$ state that leads to the nontrivial tree level behavior of $F_\pi$ in the linear sigma model.

\subsection{Explicit Symmetry Breaking and Power Counting}
In the underlying gauge theory, the chiral symmetry is explicitly broken by the quark masses.  We include this effect in the EFT by spurion analysis.  An external scalar field, $\chi$, is introduced which transforms as $\chi \rightarrow L \chi R^\dagger$.  We write down all operators constructed out of $\chi$ and $M$, and then set $\chi$ equal to a constant to break the symmetry.  For the case of degenerate quarks in the underlying theory, the spurion is set proportional to the identity matrix, $\chi \rightarrow \chi_0 1$.  Mass-split theories may be considered by different choices for the spurion v.e.v., such as in a scenario with two light quarks and $N_f - 2$ heavy quarks: $\chi \rightarrow \text{diag}(m_l, m_l, m_h,...,m_h)$.  We leave such considerations for a future work.

We now have two expansions in our EFT.  In the chiral limit, we have the standard EFT expansion, $\mathcal{L} = \sum_{\mathcal{O}} \mathcal{O}(x) / \Lambda^{d_{\mathcal{O}}-4}$ ordered in decreasing powers of $\Lambda$, where $\mathcal{O}$ is an operator constructed out of $M(x)$ and derivatives with engineering dimension $d_{\mathcal{O}}$.  $\Lambda$ is the breakdown scale of the EFT assumed to be similar in size to the lightest state excluded from the theory, $\Lambda \sim M_\rho$.  The second expansion is in powers of the spurion, which is assumed to be a small deformation.  In principle, this expansion need not be related to the EFT expansion, but rather can have its own radius of convergence.


We work with the following leading order breaking potential.
\begin{equation}
V_{SB} = - c_1 \left\langle \chi^\dagger M + \chi M^\dagger \right\rangle - c_2 \left\langle M^\dagger M \right\rangle\left\langle \chi^\dagger M + \chi M^\dagger \right\rangle - c_3 \left\langle \left( M^\dagger M \right) \left( \chi^\dagger M + M^\dagger \chi \right) \right\rangle \label{eq:Vsb}
\end{equation}
We have chosen to include terms up to $\mathcal{O}(\chi^1,M^3)$.  When we truncate the chiral expansion at first order, we find that $\chi \sim M_\pi^2$.  So as long as $M_\pi^2 / M_\rho^2$ is small, we consider it reasonable to neglect the $\mathcal{O}(\chi^2)$ terms at leading order.  In the $N_f = 8$ theory, $M_\pi^2 / M_\rho^2 \approx 1/4$ at the currently reachable distance from the chiral limit \cite{Appelquist:2016viq}.  After truncating the chiral expansion, we include all relevant and marginal operators allowed by the spurion analysis.  

We do not consider Eq.~\ref{eq:Vsb} to be the most optimal or most general leading order breaking potential.  In particular, from Eqs.~\ref{eq:minimum},\ref{eq:LO1},\ref{eq:LO2},\ref{eq:Vsb} one can show that at leading order $M_\sigma^2 \geq 3 M_\pi^2$ for this breaking potential.  This bound leads to difficulties in simultaneously fitting $M_\pi$ and $M_\sigma$.  We leave more general considerations about the breaking potential, the power counting, and the size of NLO contributions to an upcoming publication \cite{Browerprogress}.

\section{Fitting to Lattice Data} \label{sec:fits}
In this section, we test the linear sigma model defined in Eqs.~\ref{eq:LLO},\ref{eq:v0},\ref{eq:Vsb} as a description of the low energy physics of walking gauge theory by performing explicit fits to lattice data for the $N_f = 8$ theory computed by the LSD collaboration \cite{Appelquist:2016viq}.  A complete analysis of the systematic errors on this data set is not yet available.   To account for possible systematics, we have included a 3\% error added in quadrature to the reported statistical uncertainty on each data point.  Since the error analysis is not rigorous, the $\chi^2$/d.o.f. is only a qualitative measure of the quality of the fit, and the results of the fits should be considered preliminary.  For comparison, we also include fits to NLO$\chi$PT.  The relevant expressions from $\chi$PT can be found in \cite{Pich:1995bw,Gasbarro:2017fmi}.  

Let us first make some remarks about scale setting.  In a walking theory, when quarks are given large masses they are less effective at screening, and so the theory confines more rapidly than it would close to the chiral limit.  
As a result, the confinement scale is pushed towards the IR relative to the lattice scale as the quark masses are reduced.  
At some sufficiently small $m_q$, the spontaneous breaking becomes as important as the explicit breaking and the confinement scale should become less sensitive to $m_q$, but lattice calculations for the $N_f = 8$ theory are still too far from the chiral limit to observe this behavior.

Given these considerations, there are two ways one might consider setting the lattice scale.  In a mass independent scheme, it is assumed that the lattice spacing is independent of the quark masses.  This means that in a walking theory the physical confinement scale changes significantly with $m_q$.  In this scheme, one should plot and fit quantities in lattice units.  An alternate scheme is a mass dependent scheme in which one chooses some observable which is a proxy for the confinement scale and insists that the physical observable is independent of $m_q$.  The lattice spacing then acquires a significant dependence on $m_q$, so it cannot be used as a consistent scale for chiral fits.  In this scheme, quantities should be plotted and fit in units of the observable whose physical value is held fixed.  We carry out two analyses: one in lattice units assuming a mass independent scheme, and the other in units of the nucleon mass, $M_N$, assuming a mass dependent scheme.  

\subsection{Fits in Lattice Units}
\begin{figure}[t!]
	\begin{center}
		\includegraphics[width=0.7\textwidth]{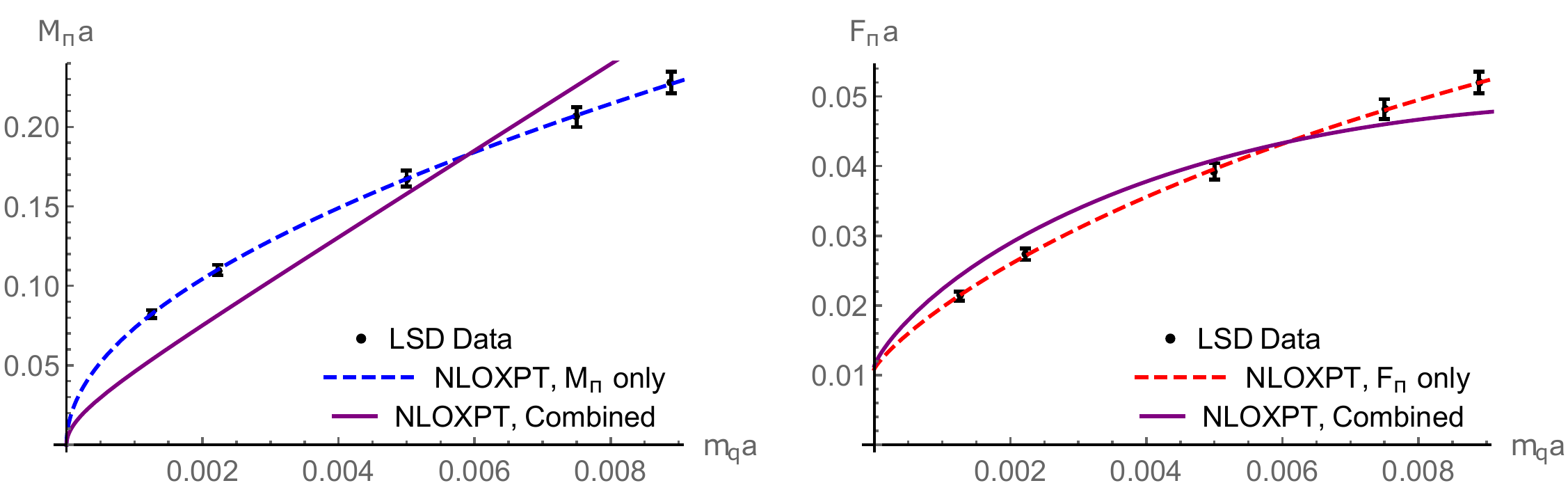}
	\end{center}
	\caption{Fits of $N_f = 8$ LSD data in lattice units to NLO$\chi$PT.  Dashed lines are fits to individual quantities.  The solid purple line is a simultaneous fit of $M_\pi$ and $F_\pi$. Fit lines are drawn for the central values of fit coefficients. Conservative 3\% error bars have been added to lattice data to account for possible systematic errors.}
	\label{fig:NLOXPT-latticeunits}
\end{figure}

\begin{figure}[t!]
	\begin{center}
		\includegraphics[width=\textwidth]{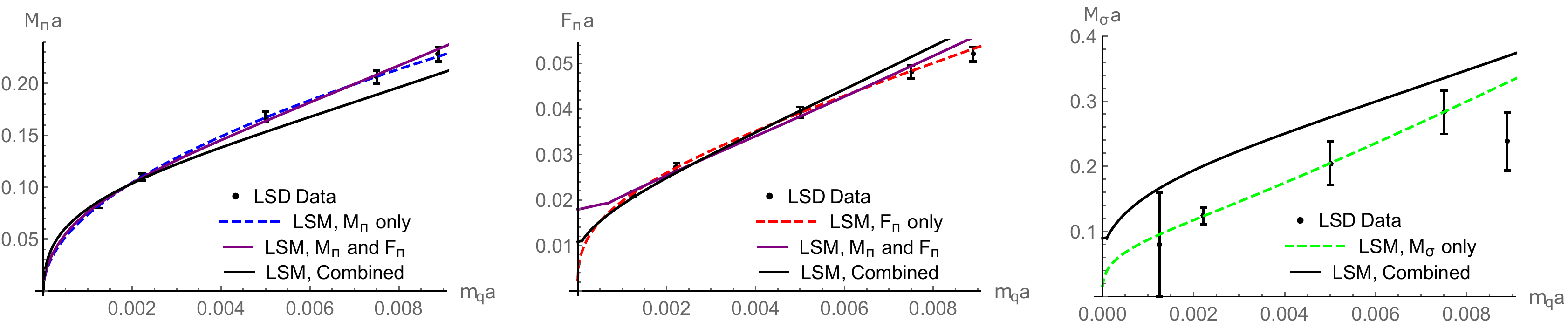}
	\end{center}
	\caption{Fits of $N_f = 8$ LSD data in lattice units to the linear sigma model.  Dashed lines are fits to individual quantities.  The solid purple line is a simultaneous fit of $M_\pi$ and $F_\pi$, and the solid black line is a simultaneous fit to all three quantities.  Fit lines are drawn for the central values of fit coefficients. Conservative 3\% error bars have been added to lattice data to account for possible systematic errors.}
	\label{fig:LSM-latticeunits}
\end{figure}
Fig.~\ref{fig:NLOXPT-latticeunits} shows the fits of NLO$\chi$PT to the data in lattice units.  While $M_\pi$ and $F_\pi$ are individually well fit by the $\chi$PT expressions, the combined fit has a $\chi^2$/d.o.f. = 296.1/10 = 29.6.  The failure of $\chi$PT to describe the Goldstone observables at this distance from the chiral limit was already pointed out in \cite{Gasbarro:2017fmi}.
Fig.~\ref{fig:LSM-latticeunits} shows the fits of the linear sigma model to the data in lattice units.  Again, each quantity -- $M_\pi$, $F_\pi$, and $M_\sigma$ -- is individually well fit by the linear sigma expressions.  The combined fit to $M_\pi$ and $F_\pi$ (excluding $M_\sigma$) also provides an acceptable fit, with a $\chi^2$/d.o.f. = 5.0/10 = 0.5.  This marks a significant improvement over chiral perturbation theory, and is one of the main results of this work.  With a linear sigma model, we are able to achieve a reduction in the $\chi^2$/d.o.f. by a factor of sixty!  
The combined fit to all three quantities has a $\chi^2$/d.o.f. = 73.5/15 = 4.9 .  In the 3rd panel of Fig.~\ref{fig:LSM-latticeunits}, one sees that the problem arises because our linear sigma model prefers a $\sigma$ mass that is slightly larger than the lattice data.

\subsection{Fits in Nucleon Units}
\begin{figure}[t!]
	\begin{center}
		\includegraphics[width=0.7\textwidth]{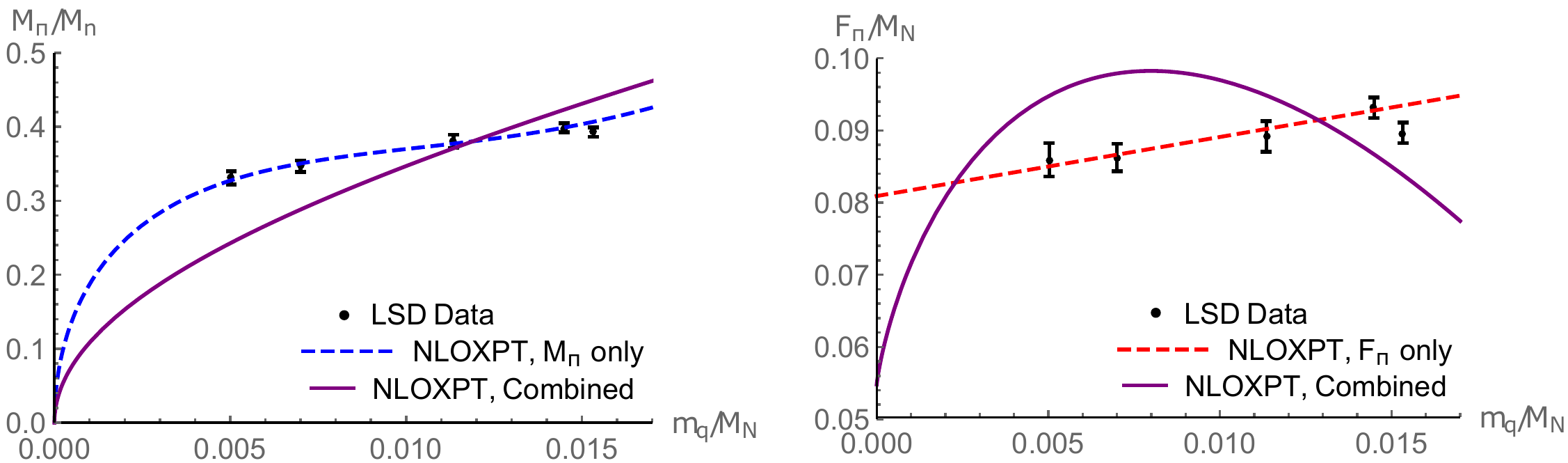}
	\end{center}
	\caption{Fits of $N_f=8$ LSD data in nucleon units to NLO$\chi$PT.  Dashed lines are fits to individual quantities.  The solid purple line is a simultaneous fit of $M_\pi$ and $F_\pi$. Fit lines are drawn for the central values of fit coefficients. Conservative 3\% error bars have been added to lattice data to account for possible systematic errors.  The heaviest mass point has been omitted from the fits.}
	\label{fig:NLOXPT-nucleonunits}
\end{figure}

\begin{figure}[t!]
	\begin{center}
		\includegraphics[width=\textwidth]{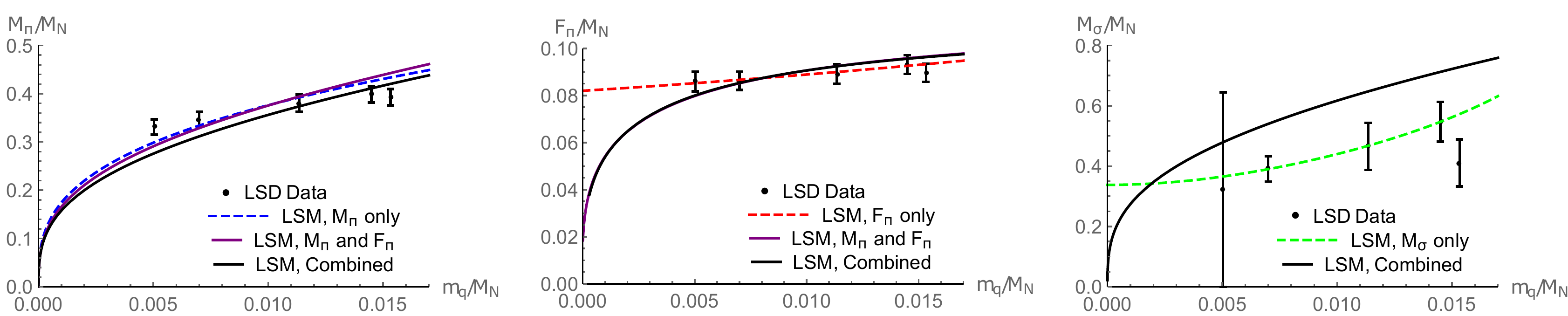}
	\end{center}
	\caption{Fits of $N_f = 8$ LSD data in nucleon units to the linear sigma model.  Dashed lines are fits to individual quantities.  The solid purple line is a simultaneous fit of $M_\pi$ and $F_\pi$, and the solid black line is a simultaneous fit to all three quantities.  Fit lines are drawn for the central values of fit coefficients. Conservative 3\% error bars have been added to lattice data to account for possible systematic errors.  The heaviest mass point has been omitted from the fits.}
	\label{fig:LSM-nucleonunits}
\end{figure}
Fig.~\ref{fig:NLOXPT-nucleonunits} shows the fits of NLO$\chi$PT to the data in nucleon units.  Again, $M_\pi$ and $F_\pi$ are individually well fit by the $\chi$PT expressions, but the combined fit has a $\chi^2$/d.o.f. = 253.8/8 = 31.7. Fig.~\ref{fig:LSM-nucleonunits} shows the fits of the linear sigma model to the data in nucleon units.  Single quantities are again individually well fit by the linear sigma model.  The combined fit to $M_\pi$ and $F_\pi$ has $\chi^2$/d.o.f. = 15.4/8 = 1.9.  The $\chi^2$/d.o.f. is larger than the fit in lattice units, but it has still been reduced by a factor of fifteen compared to NLO$\chi$PT.  Given that this is an initial analysis that does not rigorously handle systematic errors, we consider this to be an acceptable fit.  The combined fit to all three quantities yields a $\chi^2$/d.o.f. = 45.1/12 = 3.76.  The larger error is again due to the fact that the linear sigma model is favoring a heavier sigma mass than the lattice data.

Thus we find that the overall quality of fit is very similar whether the fits are performed in lattice units or nucleon units.  We remark that the $\chi^2$/d.o.f. is consistently reduced when one removes the heaviest mass point from the fit.  This suggest that the lattice data is close to the edge of the radius of convergence of the linear sigma model.  The fits would likely be improved by going to NLO in the linear sigma model, but we leave this to a future work.

\section{Conclusions and Future Work}
We have put forward a new EFT approach for describing walking gauge theories away from the chiral limit.  The model is very successful at fitting the Goldstone observables even at leading order.  In a combined fit to $M_\pi$ and $F_\pi$, we have been able to reduce the $\chi^2$/d.o.f. by an order of magnitude compared to NLO$\chi$PT.  A combined fit of $M_\pi$, $F_\pi$ and $M_\sigma$ gives a $\chi^2$/d.o.f. around 4 or 5.  The obstruction to an acceptable fit likely arises from the leading order bound $M_\sigma^2 \geq 3 M_\pi^2$ that is inherent to our breaking potential Eq.~\ref{eq:Vsb}.  We hope that a better fit will be provided by considering a more general leading order breaking potential than the simple one constructed here.  In an upcoming publication \cite{Browerprogress}, we plan to report on more general breaking potentials and power countings, estimates of the size of NLO effects, and the role of the flavored scalar $a_0$ states.

\bibliography{Lattice2017_357_GASBARRO}{}
\bibliographystyle{unsrt}

\end{document}